\documentclass[a4paper,11pt]{article}
\usepackage{epsfig}
\usepackage{amssymb}
\usepackage{amsmath}

\begin{document}
\title{LOGIC AND STRING THEORY}
\author{\Large{A. Nicolaidis} \medskip \\
Theoretical Physics Department \\
University of Thessaloniki \\
54124 Thessaloniki, Greece \\
nicolaid@auth.gr}

\maketitle

\begin{abstract}
The unification of Quantum Mechanics and General Relativity remains the primary goal of Theoretical Physics, with string theory appearing as the
only plausible unifying scheme. In the present work, in a search of the conceptual foundations of string theory, we analyze the relational logic
developed by C. S. Peirce in the late nineteenth century. The Peircean logic has the mathematical structure of a category with the relation $R_{ij}$
among two individual terms $S_i$ and $S_j$, serving as an arrow (or morphism). We introduce a realization of the corresponding categorical algebra
of compositions, which naturally gives rise to the fundamental quantum laws, thus underscoring the relational character of Quantum Mechanic.
The same relational algebra generates a number of group structures, among them $W_{\infty}$. The group $W_{\infty}$ is embodied and realized
by the matrix models, themselves closely linked with string theory. It is suggested that relational logic and in general category theory may
provide a new paradigm, within which to develop modern physical theories.
\end{abstract}

\bigskip
\bigskip
\indent The raison d' \^{e}tre of physics is to understand the
wonderful variety of nature in a unified way. A glance at the
history of physics is revealing: the unification of terrestrial
and celestial Mechanics by Newton in the 17th century; of optics
with the theories of electricity and magnetism by Maxwell in the
19th century; of space-time geometry and the theory of gravitation
by Einstein in the years 1905 to 1916; and of thermodynamics and
atomic physics through the advent of Quantum Mechanics in the
1920s \cite{c1}. The next leap in this on-going process is the
unification of the two pillars of modern physics, quantum
mechanics and general relativity. String theory, in this respect,
appears as the most promising example of a candidate unified
theory \cite{c2}.

Strings emerged in the study of strong interactions, modelling the
flux tubes between quark-antiquark pairs in hadronic collisions,
in the Regge limit, nicely described by the Veneziano amplitude
\cite{c3}, which can be reproduced from a relativistic string
theory \cite{4c}. In a similar vein, the hadronic structure
functions in the small x-Bjorken limit are most conveniently
described via colored dipoles \cite{5c}. A precise and profound
analysis of a string dual of QCD has been provided by 't Hooft
\cite{6c}. 't Hooft considered a generalization of QCD by
replacing the gauge group SU(3) by SU(N). The limit $N\rightarrow
\infty$ with $\lambda \equiv g_{YM}^{2}N$ kept fixed, leads to a
topological expansion. The leading order (in $1/N$) Feynmann
diagrams can be drawn on a planar surface and higher order
diagrams on surfaces of higher genus. In a most interesting
development an holographic analogy \cite{7c,8c} has been
established between matter or open strings on a D-brane and
gravity or closed strings in the bulk \cite{9c}. We realize that
string theory is a tantalizing rich theory, since on one hand is
connected to the dynamics of the space-time continuum, and on the
other hand the discrete modes of string vibrations represent the
totality of elementary particles.

Every single physical theory is corroborated or disproved by
experiment. The early hope of making direct contact between
experiment and string theory has long since dissipated, and there
is as yet no experimental program for finding even indirect
manifestations of underlying string degrees of freedom in nature
\cite{10c}. Particle/string theorists under these conditions
focused their attention in searching for the internal coherence
and the physical principles governing string theory. This search
is of paramount importance. While in developing general relativity
Einstein was guided by the principle of equivalence, we are
lacking a foundational principle for either string theory or
quantum mechanics \cite{c1,11c}. In the present work we suggest
that a form of logic, relational logic developed by C. S. Peirce
in the second half of the 19th century, may serve as the
conceptual foundation of quantum mechanics and string theory.

Peirce, a most original mind, made important contributions in
science, philosophy, semiotics and notably in logic, where he
invented and elaborated novel system of logical syntax and
fundamental logical concepts. The starting point is the binary
relation $S_iRS_j$ between the two 'individual terms' (subjects)
$S_j$ and $S_i$. In a short hand notation we represent this
relation by $R_{ij}$. Relations may be composed: whenever we have
relations of the form $R_{ij}$, $R_{jl}$, a third transitive
relation $R_{il}$ emerges following the rule \cite{12c,13c}
\begin{equation}
 R_{ij}R_{kl}=\delta_{jk}R_{il}
\label{1}
\end{equation}
In ordinary logic the individual subject is the starting point and it is defined as a member of a set. Peirce, in an original move, considered the individual as the aggregate of all its relations
\begin{equation}
 S_i =\sum_{j}R_{ij}.
\label{2}
\end{equation}
It is easy to verify that the individual $S_i$ thus defined is an eigenstate of the $R_{ii}$ relation
\begin{equation}
 R_{ii}S_i =S_i.
\label{3}
\end{equation}
The relations $R_{ii}$ are idempotent
\begin{equation}
 R_{ii}^{2}=R_{ii}
\label{4}
\end{equation}
and they span the identity
\begin{equation}
 \sum_{i}R_{ii}=\textbf{1}
\label{5}
\end{equation}
The Peircean logical structure bears great resemblance to category
theory, a remarkably rich branch of mathematics developed by
Eilenberg and Maclane in 1945 \cite{14c}. In categories the
concept of transformation (transition, map, morphism or arrow)
enjoys an autonomous, primary and irreducible role. A category
\cite{15c} consists of objects A, B, C,... and arrows (morphisms)
f, g, h,... . Each arrow f is assigned an object A as domain and
an object B as codomain, indicated by writing $f:A \rightarrow B$.
If g is an arrow $g:B \rightarrow C$ with domain B, the codomain
of f, then f and g can be ``composed'' to give an arrow $gof:A
\rightarrow C$. The composition obeys the associative law
$ho(gof)=(hog)of$. For each object A there is an arrow $1_{A}:A
\rightarrow A$ called the identity arrow of A. The analogy with
the relational logic of Peirce is evident, $R_{ij}$ stands as an
arrow, the composition rule is manifested in eq. (\ref{1}) and the
identity arrow for $A \equiv S_i$ is $R_{ii}$. There is an
important literature on possible ways the category notions can be
applied to physics; specifically to quantising space-time
\cite{16c}, attaching a formal language to a physical system
\cite{17c}, studying topological quantum field theories
\cite{18c,19c}.

$R_{ij}$ may receive multiple interpretations: as a transition
from the j state to the i state, as a measurement process that
rejects all impinging systems except those in the state j and
permits only systems in the state i to emerge from the apparatus,
as a transformation replacing the j state by the i state. We
proceed to a representation of $R_{ij}$
\begin{equation}
 R_{ij}=\left.| r_i \right\rangle \left\langle r_j\right.|
\label{6}
\end{equation}
where state $\left\langle r_i\right.|$ is the dual of the state $\left.| r_i \right\rangle$ and they obey the orthonormal condition
\begin{equation}
 \left\langle r_i\right.\left.| r_j \right\rangle=\delta_{ij}
\label{7}
\end{equation}
It is immediately seen that our representation satisfies the
composition rule eq. (\ref{1}). The completeness, eq.(\ref{5}),
takes the form
\begin{equation}
 \sum_{n}\left.| r_i \right\rangle \left\langle r_i\right.|=\textbf{1}
\label{8}
\end{equation}
All relations remain satisfied if we replace the state $\left.| r_i \right\rangle$ by $\left.| \varrho_i \right\rangle$, where
\begin{equation}
 \left.| \varrho_i \right\rangle=\frac{1}{\sqrt{N}}\sum_{n}\left.| r_i \right\rangle \left\langle r_n\right.|
\label{9}
\end{equation}
with N the number of states. Thus we verify Peirce's suggestion,
eq. (\ref{2}), and the state $\left.| r_i \right\rangle$ is
derived as the sum of all its interactions with the other states.
$R_{ij}$ acts as a projection, transferring from one r state to
another r state
\begin{equation}
 R_{ij}\left.| r_k \right\rangle=\delta_{jk}\left.| r_i \right\rangle.
\label{10}
\end{equation}
We may think also of another property characterizing our states and define a corresponding operator
\begin{equation}
 Q_{ij}=\left.| q_i \right\rangle \left\langle q_j\right.|
\label{11}
\end{equation}
with
\begin{equation}
 Q_{ij}\left.| q_k \right\rangle=\delta_{jk}\left.| q_i \right\rangle.
\label{12}
\end{equation}
and
\begin{equation}
\sum_{n}\left.| q_i \right\rangle \left\langle q_i\right.|=\textbf{1}.
\label{13}
\end{equation}
Successive measurements of the q-ness and r-ness of the states is provided by the operator
\begin{equation}
 R_{ij}Q_{kl}=\left.| r_i \right\rangle \left\langle r_j\right.\left.| q_k \right\rangle \left\langle q_l\right.|=
\left\langle r_j\right.\left.| q_k \right\rangle S_{il}
\label{14}
\end{equation}
with
\begin{equation}
 S_{il}=\left.| r_i \right\rangle \left\langle q_l\right.|.
\label{15}
\end{equation}
Considering the matrix elements of an operator A as $A_{nm}=\left\langle r_n\right.|A\left.| r_m \right\rangle$ we find for the trace
\begin{equation}
 Tr(S_{il})=\sum_{n}\left\langle r_n\right.| S_{il} \left.| r_n \right\rangle=\left\langle q_l\right. \left.| r_i \right\rangle.
\label{16}
\end{equation}
From the above relation we deduce
\begin{equation}
 Tr(R_{ij})=\delta_{ij}.
\label{17}
\end{equation}
Any operator can be expressed as a linear superposition of the $R{ij}$
\begin{equation}
 A=\sum_{i,j}A_{ij}R_{ij}
\label{18}
\end{equation}
with
\begin{equation}
 A_{ij}=Tr(AR_{ji}).
\label{19}
\end{equation}
The individual states can be redefined
\begin{eqnarray}
 \left.| r_i \right\rangle \longrightarrow e^{i\varphi_i}\left.| r_i \right\rangle \label{20}\\
 \left.| q_i \right\rangle \longrightarrow e^{i\theta_i}\left.| q_i \right\rangle
\label{21}
\end{eqnarray}
without affecting the corresponding composition laws. However the overlap number $\left\langle r_i\right. \left.| q_j \right\rangle$ changes
 and therefore we need an invariant formulation for the transition $\left.| r_i \right\rangle \rightarrow \left.| q_j \right\rangle$.
 This is provided by the trace of the closed operation $R_{ii}Q_{jj}R_{ii}$
\begin{equation}
 Tr(R_{ii}Q_{jj}R_{ii})\equiv p(q_j,r_i)= \left.|\left\langle r_i\right. \left.| q_j \right\rangle\right.|^2.
\label{22}
\end{equation}
The completeness relation, eq. (\ref{13}), guarantees that
$p(q_j,r_i)$ may assume the role of a probability since
\begin{equation}
 \sum_{j}p(q_j,r_i)=1.
\label{23}
\end{equation}
We discover that starting from the relational logic of Peirce we
obtain all the essential laws of Quantum Mechanics. Our derivation
underlines the outmost relational nature of Quantum Mechanics and
goes in parallel with the analysis of the quantum algebra of
microscopic measurement presented by Schwinger \cite{20c}.

Further insights are obtained if we consider the simplified case of only two states (i=1,2). We define
\begin{equation}
 R_z=\frac{1}{2}(R_{11}-R_{22})
\label{24}
\end{equation}
and
\begin{equation}
 R_+=R_{12}\;\;\;\;\;R_-=R_{21}.
\label{25}
\end{equation}
These operators satisfy the SU(2) commutation relations
\begin{equation}
 [R_z,R_{\pm}]=\pm R_{\pm}\;\;\;\;[R_+,R_-]=2R_z
\label{26}
\end{equation}
and the quadratic Casimir operator
\begin{equation}
 R^2=R_{z}^2+\frac{1}{2}(R_+R_-+R_-R_+)
\label{27}
\end{equation}
can be written as
\begin{equation}
 R^2=\frac{1}{2}(\frac{1}{2}+1)\textbf{1}.
\label{28}
\end{equation}
The underlying dynamics is analogous to an ``angular momentum
$1/2$ particle'' and the SU(2) algebra is realized in a way
reminiscent of the Schwinger scheme \cite{21c,22c}. A matrix
representation of $R_{ij}$, for the two-states case,
  is provided by
\begin{eqnarray}
 R_{11}=
\left(
\begin{array}{cc}
 1& 0 \\
0 & 0
 \end{array}
\right) \;\;\;\;\;\;
R_{22}=
\left(
\begin{array}{cc}
 0& 0 \\
0 & 1
 \end{array}
\right) \\ \bigskip
 R_{12}=
\left(
\begin{array}{cc}
 0& 1 \\
0 & 0
 \end{array}
\right) \;\;\;\;\;\;
R_{21}=
\left(
\begin{array}{cc}
 0& 0 \\
1 & 0
 \end{array}
\right)
\label{29}
\end{eqnarray}
The matrices
\begin{eqnarray}
 exp(sR_{12})=
\left(
\begin{array}{cc}
 1& s \\
0 & 1
 \end{array}
\right) \label{30} \\ \bigskip
exp(tR_{21})=
\left(
\begin{array}{cc}
 1& 0 \\
t & 1
 \end{array}
\right)
\label{31}
\end{eqnarray}
perform shear transformations in a two-dimensional space
\cite{23c}, while the matrix
\begin{equation}
 exp[\eta(R_{11}-R_{22})]=
\left(
\begin{array}{cc}
 e^{\eta} & 0 \\
0 & e^{-\eta}
 \end{array}
\right)
\label{32}
\end{equation}
generates squeeze transformations.

\begin{figure}[!h]
\begin{center}
\includegraphics[scale=.6]{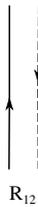}
\end{center}
\caption{The relation $R_{12}$.  Solid (dashed) line stands for
the state 1 (2).  A downward (upward) arrow is attached to an
impinging (emerging) state.} \label{fig:1a}
\end{figure}

\begin{figure}[!h]
\begin{center}
\includegraphics[scale=.6]{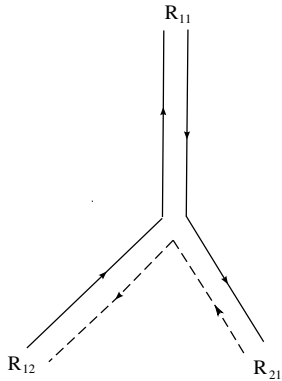}
\end{center}
\caption{Pictorial representation of the composition rule
$R_{12}R_{21} = R_{11}$} \label{fig:1b}
\end{figure}

\begin{figure}[!h]
\begin{center}
\includegraphics[scale=.6]{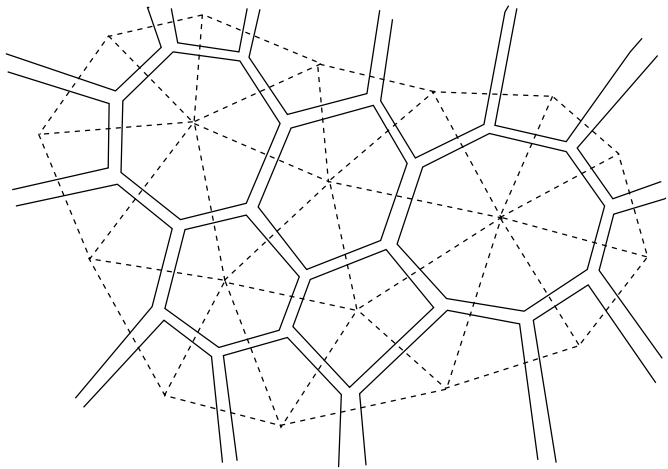}
\end{center}
\caption{Random partition of a surface. Each triangle (dashed
lines) is dual to a cubic vertex.} \label{fig:2}
\end{figure}

For the general case of N available states the $R_{ij}$ satisfy
the $W_{\infty}$ algebra
\begin{equation}
 [R_{ij},R_{kl}]=\delta_{jk}R_{il}-\delta_{li}R_{kj}.
\label{33}
\end{equation}
The $W_{\infty}$ algebras are bosonic extensions of the Virasoro
algebra, containing generating currents of higher conformal-spin,
in addition to the spin-2 stress tensor of Virasoro (for a review
see \cite{24c}). They are linked to the area-preserving
diffeomorphisms of two dimensional surfaces \cite{25c,26c}.
$W_{\infty}$ symmetries are exhibited by a number of systems,
among them, $QCD_2$ \cite{27c,28c}, gravity in two-dimensions
\cite{29c}, bosonic string in four-dimensional Minkowski space
\cite{30c}. We may proceed to a pictorial representation of the
operation $R_{ij}$. Each distinct state i is represented by a
specific line (solid, dashed,...), with a downward (upward) arrow
attached to the annihilated (created) state. In this sense we
picture $R_{12}$ by a double line, fig \ref{fig:1a}, while the
composition rule,
for example $R_{12}R_{21}=R_{11}$, is represented by the diagram of fig. \ref{fig:1b}.
The similarity with string theory, string joining and string splitting, is obvious.
The ``cubic-string'' interaction may be repeated an indefinite number of times, with vertices
connected together and giving rise to different forms of polygons (see fig. \ref{fig:2}).
These types of structures can be generated by a random matrix
model \cite{10c}
\begin{equation}
 Z=\int [dM] exp\{-N tr(\frac{1}{2}M^2+gM^3)\}
\label{34}
\end{equation}
where M are $N\times N$ random matrices. A perturbative expansion of this integral leads to
't Hooft-type Feynman diagrams with cubic vertices. Each such diagram specifies a unique surface topology,
with faces arbitrary n-gons. The corresponding dual lattice has n lines meeting at a point but the faces are triangles.
The result is a triangulated Riemann surface (fig. \ref{fig:2}). An expansion of Z in inverse powers of
N is equivalent to a topological expansion, selecting diagrams of specific genus h
\begin{equation}
 Z=\sum_{h=0}^{\infty}Z_h(g)N^{2-2h}.
\label{35}
\end{equation}
As g is increased successive contributions $Z_h$ diverge at the
same critical value $g=g_c$. The partition function can be
reorganized into
\begin{equation}
 Z=\sum_{h}F_h g_s^{2h-2}
\label{36}
\end{equation}
where the ``renormalized'' string coupling $g_s$ is given by
\begin{equation}
 g_s=\frac{1}{N(g-g_s)^{\frac{2-\gamma}{2}}}
\label{37}
\end{equation}
with $\gamma$ the critical exponent. The continuum two dimensional
string theory is obtained in the double scaling limit $N
\rightarrow \infty$, $g \rightarrow g_c$ with $g_s$ kept fixed
\cite{31c}.

Modern physics is marked by two impressive theoretical
constructions, quantum mechanics and string theory. Each of them
is an elaborate and detailed theory providing understanding or
insights to a host of different problems. Yet, we are lacking a
conceptual foundation for these theories. In the present work we
have indicated that a form of logic, relational logic developed by
C. S. Peirce, may serve as the foundation of both quantum
mechanics and string theory. The starting point is that the
concept of relation is an irreducible basic datum. All other terms
or objects are defined in terms of relations, transformations,
morphisms, arrows, structures. Usually we adhere to mathematical
considerations derived within set theory. A set is deprived of any
structure, being a plurality of structureless individuals,
qualified only by membership (or non membership). Accordingly a
set-theoretic enterprise is analytic, atomistic, arithmetic. On
the other hand a relational or categorical formulation is bound to
be synthetic, holistic, geometric. It appears that quantum theory,
string theory and eventually the physical theories to come, are
better conceived, analyzed and comprehended within a new paradigm
inspired by relational and categorical principles.\newline
\medskip

\noindent \textbf{Acknowledgement}: Part of the present work was presented during the workshop on ``Relational Ontology'',
held at the Academy of Athens (October 14 - 17 2005) and organized by the Templeton Foundation.

\end{document}